\documentclass[showpacs,preprintnumbers,amsmath,amssymb,nofootinbib]{revtex4}

\usepackage[pdftex]{graphicx}
\usepackage{amsmath}
\usepackage{amssymb}
\usepackage{epstopdf}

\usepackage{dcolumn}
\usepackage{bm}

\newcommand{\be}{\begin{equation}}
\newcommand{\ee}{\end{equation}}
\newcommand{\bear}{\begin{eqnarray}}
\newcommand{\eear}{\end{eqnarray}}
\newcommand{\ba}{\begin{array}}
\newcommand{\ea}{\end{array}}

\newskip\humongous \humongous=0pt plus 1000pt minus 1000pt

\newif\ifdtup

\def\oldreffmt#1{\rlap{[#1]} \hbox to 2\parindent{}}

\def\figfmt#1{\rlap{Figure {#1}} \hbox to 1in{}}

\def\beq{\begin{equation}}
\def\eeq{\end{equation}}
\def\bea{\begin{eqnarray}}
\def\eea{\end{eqnarray}}

\def\bq{\begin{quote}}
\def\eq{\end{quote}}

\relax

\newdimen\tdim
\tdim=\unitlength
\def\bar{\overline}

\begin{document}
\setcounter{page}{0}

\begin{flushright}
ANL-HEP-PR-10-11\\ 
NUHEP-TH/10-03\\
UCI-TR-2010-07\\
\end{flushright}

\title{\LARGE Beautiful Mirrors at the LHC}

\author{ { Kunal Kumar$^{a,b}$, William Shepherd$^{b,c}$, Tim M.P. Tait$^{a,b,c}$, 
and Roberto Vega-Morales$^{a,b}$}\\[0.25cm]
\normalsize{$^{a}$ HEP Division, Argonne National Lab, Argonne IL 60439.\\
\normalsize{$^{b}$ Northwestern University, 2145 Sheridan Road, Evanston, IL 60208.\\
\normalsize{$^{c}$ Department of Physics and Astronomy, University of California, Irvine, CA 92697.}}}}

\begin{abstract}
We explore the ``Beautiful Mirrors" model, which aims to explain the measured value of
$A^b_{FB}$, discrepant at the $2.9\sigma$ level.  This scenario introduces
vector-like quarks which mix with the bottom, subtly affecting its coupling to the $Z$.
The spectrum of the new particles consists of two bottom-like quarks and a charge
$-4/3$ quark, all of which have electroweak interactions with the third generation.
We explore the phenomenology and discovery reach for these new particles at the LHC,
exploring single mirror quark production modes whose rates are proportional to the
same mixing parameters which resolve the $A_{FB}^b$ anomaly.  We find that for mirror
quark masses $\lesssim 500$~GeV, a 14 TeV LHC with 300~${\rm fb}^{-1}$ is required
to reasonably establish the scenario and extract the relevant mixing parameters.
\end{abstract}

\maketitle
\thispagestyle{empty} 

\section{Introduction}
\label{sec:intro}

The primary mission of the Large Hadron Collider (LHC) is to seek evidence for the
breakdown of the Standard Model (SM) \cite{Morrissey:2009tf}.  
For the most part, the SM with a light Higgs provides a 
very accurate description of the observed data coming from a wide variety of experiments.  While
deviations from the SM have come (and mostly gone), most disappear as statistics and
experimental precision increases and theoretical inputs improve.  The agreement between the
SM predictions and experiment is unprecedented, particularly in the arena of precision 
electroweak measurements, many of which have per mil level uncertainties \cite{LEP:2009}.

However, there is one notable exception.  The forward-backward asymmetry of the bottom
quark ($A^{b}_{FB}$) shows roughly a $2.9\sigma$ 
deviation\footnote{It is interesting that recent Tevatron
measurements also show an unexpected asymmetry in top quark pair 
production \cite{:2007qb}, though existing proposed new physics
explanations do not typically correlate this
with any particular effect on $A^{b}_{FB}$ \cite{Antunano:2007da}.} from the value predicted 
by a best fit to precision data within the SM
 \cite{LEP:2009}.  While not in itself very significant,
this deviation has persisted for more than a decade and may be a guide to what the LHC
could find.  $A^{b}_{FB}$ further plays an interesting role in the global fit to precision data, which
in the context of the SM provides the indirect constraints on the 
Higgs mass \cite{Chanowitz:2002cd}.  Indeed, the poor
fit to $A^{b}_{FB}$ can be understood as a tension in the preferred value of $m_h$ between
the leptonic observables, which prefer $m_h \sim 50$~GeV and $A^{b}_{FB}$ which
prefers values closer to $\sim 1$~TeV.  The fit has settled into an ``unhappy" middle ground
between the two, favoring the other measurements at the cost of disagreeing with
the observed  $A^{b}_{FB}$.  As a result, if one simply ignores  $A^{b}_{FB}$, one gets a
fit to $m_h$ which is marginally at odds with the direct search bound from 
LEP-II \cite{Ludwig:2006ar}.

Broadly defined, there are three attitudes one can take toward $A^{b}_{FB}$ and the
precision data:
\begin{itemize}
\item One can assume $A^{b}_{FB}$ is a statistical (or unaccounted for
systematic) effect and that reasonable variations of the other measured precision observables
explain the tension between the SM fit to the Higgs mass and direct searches.
\item One can consider the possibility that $A^{b}_{FB}$ itself does not reflect the
presence of new physics, but accidently makes the SM fit to $m_h$ more palatable than it
would otherwise have been.
In this case, one can invoke new physics
contributions to the Peskin and Takeuchi $T$ parameter \cite{Peskin:1991sw}
which may reconcile the indirect bounds on $m_h$ with the direct search limits 
(for a few examples, see \cite{Peskin:2001rw}).
\item One can take the attitude that the bottom quark couplings to the $Z$ boson may themselves
reflect the presence of new 
physics \cite{Haber:1999zh,Choudhury:2001hs,Morrissey:2003sc}.
\end{itemize}
In this article, we will take the last approach, and explore the consequences of one
particular model of this kind, the ``Beautiful Mirrors" model \cite{Choudhury:2001hs}, which
introduces new physics to produce the observed
anomaly in $A^{b}_{FB}$.

The Beautiful Mirrors model works by introducing a new set of vector-like (or ``mirror")
quarks, which
mix with the bottom quark, adjusting its coupling to the $Z$.  Vector-like quarks are
chosen so that gauge anomalies are trivially evaded, and the requirement that
there be no source of EWSB other than the SM Higgs (motivated to avoid tree-level contributions
to the oblique electroweak parameters) restricts the $SU(2)$ representations of the mirror
quarks to singlets, triplets, and doublets.  In \cite{Choudhury:2001hs}, two versions of the
doublet model were explored.
The desired shift in the $Z$ couplings to bottom quarks may be effected
for mirror quark quantum numbers under ($SU(3)$, $SU(2)$, $U(1)$) given by
$(3,2,1/6)$ or $(3,2,-5/6)$.  The first option looks like a vector-like fourth SM generation,
and requires mirror quark masses $\lesssim 400$~GeV and SM Higgs mass
$m_h \gtrsim 300$~GeV in order to fit the LEP data.  Its detailed phenomenology was
explored in Ref.~\cite{Morrissey:2003sc}.  The null results for direct searches for the mirror
quarks \cite{cdf-tpsearch}
have severely restricted the parameter space of this model, leading us to consider the more
exotic representation $(3,2,-5/6)$, which contains a bottom-like mirror quark $\omega$ and
its electroweak partner, an electrically charged $-4/3$ quark, $\chi$.  The precision
data favors the masses for these ``exotic mirrors" to be $\gtrsim 500$~GeV
\cite{Choudhury:2001hs}, making them
perfect targets for a discovery at the LHC.

In this paper we explore the phenomenological consequences of the Exotic 
Mirrors model at the LHC.  We begin in Section~\ref{sec:zbb} by revisiting the
target $Z$-$b$-$\bar{b}$ couplings, which helps pin down the amount of mixing 
required when we discuss the beautiful mirrors model itself in Section~\ref{sec:model}.
LHC signals and strategies to establish a given signal as arising from the mirror quark
solution to the $A^{b}_{FB}$ puzzle are presented in Section~\ref{sec:pheno}.  We conclude
in Section~\ref{sec:conclusion}.

\section{ {\boldmath $Z$} Boson Couplings to Bottom Quarks}
\label{sec:zbb}

\begin{figure}
\begin{minipage}[b]{0.5\linewidth}
\centering
\includegraphics[width=8.6 cm]{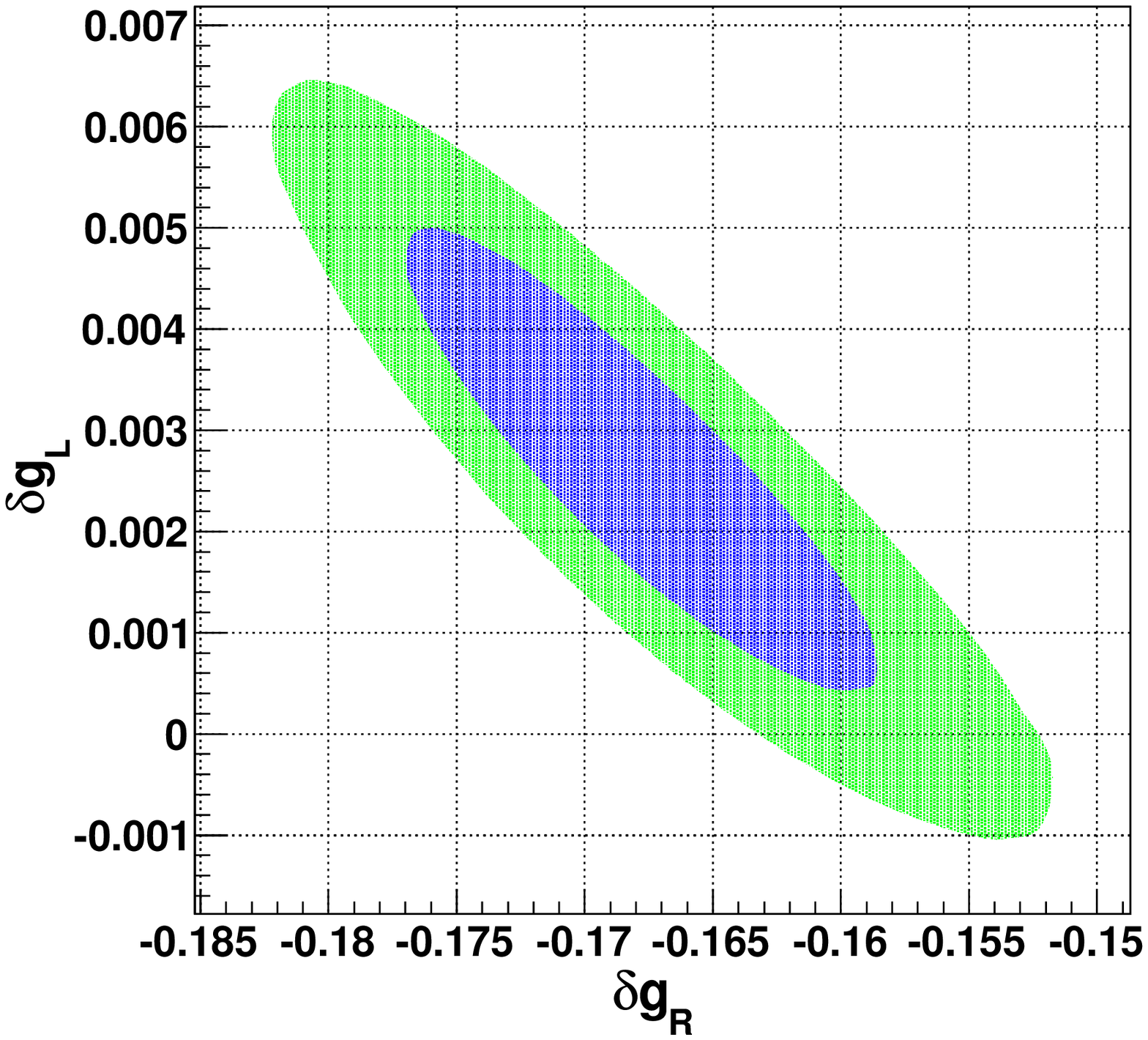} 
\end{minipage}
\hspace{-0.2 cm}
\begin{minipage}[b]{0.5\linewidth}
\centering
\includegraphics[width=8.6 cm]{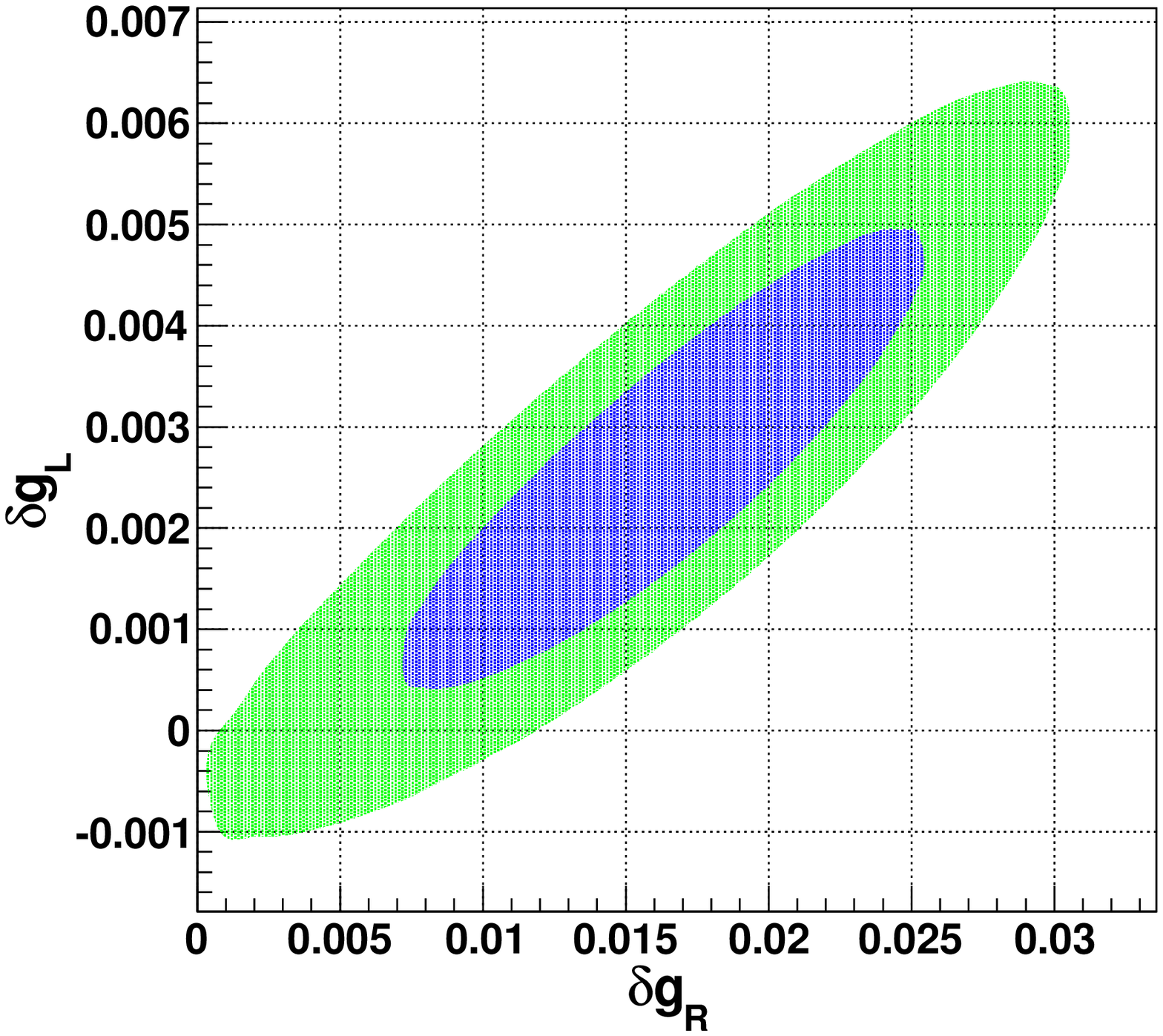} 
\end{minipage}
\caption{\label{fig:deltag} The regions in the $Z$-$b$-$\bar{b}$ coupling parameter 
space favored by EW precision data.  The inner (outer) shaded regions correspond to $1\sigma$ 
($2\sigma$) agreement with the best fit shifts in the left- and right-handed  couplings.}
\end{figure}

In this section, we examine the ranges of values of the $Z$-$b$-$\bar{b}$ couplings
consistent with precision electroweak data.
Modifications to bottom couplings must be applied subtly.  While $A^{b}_{FB}$ is discrepant
as described above,
the branching ratio of $Z$ bosons decaying into $b \bar{b}$ (which is usually reported as
a ratio between the decay to bottom quarks and into all hadrons, 
$R_b \equiv \Gamma ( Z \rightarrow b \bar{b} ) / \Gamma ( Z \rightarrow \mathrm{hadrons} )$)
shows no large deviation  \cite{LEP:2009}.  In addition, as discussed in \cite{Choudhury:2001hs},
there is data from off of the $Z$-pole which, while less precise than the $Z$-pole measurements,
implies important constraints on the signs of the couplings.  In particular, the off-pole
data requires that the left-handed interaction be close to the SM value, but does not
restrict the sign of the right-handed value.  Since together $A^{b}_{FB}$ and $R_b$ restrict
the magnitude of the couplings, the allowed space of couplings lies within two disjoint
regions of parameter space.

To explore the allowed regions of coupling space, we allow shifts in the left- and
right-handed $Z$-$b$-$\bar{b}$ interactions by $\delta g_L$ and $\delta g_R$, respectively.
We include these parameters in a global fit to the precision data, 
including the Tevatron measurements of
the top and $W$ masses \cite{:2009ec}.  We marginalize over
 $\alpha_{EM}$, $\alpha_{S}$, $m_{t}$, $m_{Z}$, and $m_{h}$, in particular allowing $m_h$
 to take any value consistent with the direct search limit from LEP-II \cite{Ludwig:2006ar}.
We assume there are no large additional contributions 
to the oblique parameters $S$ and $T$ beyond those which result from varying the top and
Higgs masses\footnote{Note that the assumption of no large additional contributions to
the $S$ and $T$ parameters is consistent with the exotic mirror scenario, but {\em not} with
standard mirror quarks.}.
The results of the fit are presented in Figure~\ref{fig:deltag}, which indicates
that the data favors small ($\sim 10^{-3}$) corrections to the left-handed coupling
and more large (either $\sim + 10^{-2}$ or $\sim - 0.2$) shifts in $\delta g_R$.  The values
of $\delta g_R$ and $\delta g_L$ should be highly correlated with one another, in order
to result in the necessary correction to $A_{FB}^b$, while the maximum and minimum changes are
set by $R_b$.

\section{Beautiful Mirrors}
\label{sec:model}

The exotic beautiful mirrors model extends the Standard Model by introducing 
two sets of vector-like quarks, $\Psi_{L,R}$ with quantum numbers $(3,2,-5/6)$
and $\xi_{L,R}$ with quantum numbers $(3,1,-1/3)$.  In terms of its $SU(2)$
components, $\Psi$ decomposes as,
\bea
 \Psi_{L,R} &=&
 \left(\begin{matrix}\omega_{L,R} \cr \chi_{L,R}\end{matrix}\right) \\
 \nonumber
\eea
where $\omega$ is a charge $-1/3$ quark and $\chi$ has charge $-4/3$.
Introducing vector-like quarks allows for new flavor mixing, which we will
ultimately invoke to explain the measured value of $A_{FB}^b$.
Among other effects, this mixing can lead to right handed $W$ couplings and tree level
flavor changing interactions with the $Z$ and Higgs.

We assume for simplicity that the exotic quarks only couple to the third generation 
SM quarks, as $Z$ couplings to the two light generations appear to agree with SM predictions 
and any corrections are thus constrained to be small.  Allowing for substantial mixing between the mirror quarks and the two lighter SM generations will generate tree level FCNC interactions which can contribute to $b \rightarrow s \gamma$ \cite{Bhattacharyya:1994gu} which is highly constrained.

In addition, mixing with the light quarks leads to interactions of the type $Z$-$b$-$s$, $Z$-$b$-$d$, and $Z$-$s$-$d$,  as well as one loop box diagrams (with the mirror quark running in the loop), contributing to $B$-$\bar{B}$ \cite{Maalampi:1987gu} and possibly $K$-$\bar{K}$ and $D$-$\bar{D}$ mixing \cite{Yanir:2002au}, all of which lead to tight constraints. These interactions are additionally constrained by rare decay processes of the strange and bottom
mesons \cite{Branco:1993gu}, as well as $B$ and $K$ meson decays such as $B \rightarrow \ell^+ \ell^- X$, $B \rightarrow J/\psi K_s$ and $K \rightarrow \pi \nu \bar{\nu}$ \cite{Barenboim:2001gu}.

That said, provided the mixing is small enough,
the presence of such mixing between the mirror quarks
and the first- and second-generation fermions (perhaps motivated by minimal flavor violation
\cite{D'Ambrosio:2002ex}) 
would not much affect the parameter space or resulting phenomenology.
The choice of exotic mirrors (as opposed to the Standard Mirror gauge assignment)
induces no right-handed $W$-$t$-$b$ interaction, evading potentially strong bounds 
again coming from $b \rightarrow s \gamma$ \cite{Larios:1999au}.

\subsection{Mixing and the Mass Eigenstates}

Electroweak symmetry breaking (EWSB) occurs as in the SM through the 
vacuum expectation value of a Higgs scalar, $\Phi$.  
We assume that the SM Higgs is the only source of
EWSB, and write down the complete set of interactions between the mirror quarks
and the third generation SM quarks, as allowed by $SU(3) \times SU(2) \times U(1)$ gauge
invariance.  In addition, the vector quarks have Dirac masses, whose magnitudes 
are not dictated by EWSB, but we will assume are at the $\sim $~TeV scale.  Such mass
terms are protected by chiral symmetries, and thus technically natural in the sense
of 't Hooft \cite{'tHooft:1980xb}.  The complete set of Yukawa interactions and masses involving the mirror
quarks are,
\bea
{\mathcal L}_{mass} = -y_{1}\overline{Q}^{\prime}_{L}\Phi b^{\prime}_{R}  
		-y_{R}\overline{\Psi}^{\prime}_{L}\tilde{\Phi}b^{\prime}_{R}
		-y_{L}\overline{Q}^{\prime}_{L}\Phi\xi^{\prime}_{R}
		-y_{5}\overline{\Psi}^{\prime}_{L}\tilde{\Phi}\xi^{\prime}_{R}
		-M_{2}\overline{\Psi}^{\prime}_{L}\Psi^{\prime}_{R}
		-M_{3}\overline{\xi}^{\prime}_{L}\xi^{\prime}_{R}
		+ h.c.
\label{eq:lmass}		
\eea
where the primed fields refer to gauge (as opposed to mass) eigenstates and
$Q^\prime_L$ refers to the third generation quark doublet and $b^\prime_R$
is the third generation down-type singlet.

After symmetry breaking, the couplings are most transparent in the unitary gauge, 
$\Phi=\frac{1}{\sqrt{2}}\left( 0 ~~ v+h \right)^T$, where $v \sim 174$~GeV is the EWSB
vacuum expectation value and $h$ is the Higgs boson.
The mass and mixing terms of the
Lagrangian may be written in matrix form,
\bea
{\mathcal L}_{mass} = - \mathbf{\overline{d}^{\prime}_{L}}
\left(\mathbf{M_{d}} + \frac{h}{v} ~ \mathbf{N_{d}} \right)  
\mathbf{{d}^{\prime}_{R}}
		+ h.c.
\eea
where $\mathbf{d^{\prime}_{L,R}} =(b_{L,R}, \omega_{L,R}, \xi_{L,R})$ are vectors in
flavor space. 
$\mathbf{M_d}$ is the bottom sector mass matrix,
\bea
	\mathbf{M_d} =
	\left( \begin{array}{ccc}
	Y_1       &      0     &     Y_L \\ \\
	Y_R      &   M_2   &      Y_5 \\ \\
	0            &      0     &     M_3 
	\end{array} \right)
\eea
where $Y_{i}=y_{i}v/\sqrt{2}$. $\mathbf{N_{d}}/v$ is the coupling matrix between the real 
Higgs and the down type quarks,
\bea
	\mathbf{N_d} =
	\left( \begin{array}{ccc}
	Y_1       &      0     &    Y_L \\ \\
	Y_R      &     0   &    Y_5 \\ \\
	0            &      0     &    0 
	\end{array} \right) ~.
\eea

To diagonalize the mass matrix we rotate by unitary matrices 
$\mathbf{U_d}$ and $\mathbf{W_d}$ 
which transform the left- and right-handed gauge eigenstates into the corresponding mass 
eigenstates (denoted by unprimed vectors in
flavor space, $\mathbf{d_{L,R}}$). We parametrize these matrices
\bea
	\mathbf{U_d} =
	\left( \begin{array}{ccc}
	c^{L}_{12} c^L_{13}      &    s^L_{12}c^L_{13} & s^L_{13} \\ \\
	-s^L_{12}c^L_{23} - c^L_{12}s^L_{23}s^L_{13}   &   c^L_{12}c^L_{23} - s^L_{12}s^L_{23}s^L_{13}   &   s^L_{23}c^L_{13} \\ \\
	s^L_{12}s^L_{23} - c^L_{12}c^L_{23}s^{L}_{13} & -c^L_{12}s^L_{23} - s^L_{12}c^L_{23}s^L_{13} & c^L_{23}c^L_{13}   
	\end{array} \right)
	\label{eq:parameterization}
\eea
where $c^L_{12} \equiv \cos \theta^L_{12}$ and so on, and with an analogous expression
for $\mathbf{W_d}$ with $\theta^L_{ij} \rightarrow \theta^R_{ij}$.  We have set potential phases to
zero for simplicity; their inclusion will complicate the analysis slightly
but are not expected to shed much
light on the $A_{FB}^b$ puzzle.
These matrices transform the gauge eigenstates to mass eigenstates,
\bea
\mathbf{d^{\prime}_{L}} & = & \mathbf{U_{d}}  ~ \mathbf{d_{L}} ~,\nonumber \\
\mathbf{d^{\prime}_{R}} & = & \mathbf{W_{d}} ~ \mathbf{d_{R}} ~.
\eea
The requirement that these transformations produce the mass eigenbasis requires
\bea
\mathbf{ U_d}^\dagger \mathbf{ M_d} \mathbf{W_d} = 
\left(
\begin{array}{ccc}
	m_1 &      0     &    0 \\ \\
	0       &  m_2   &    0 \\ \\
	0       &      0     &    m_3 
	\end{array}
\right) ~.
\eea
For values of the mixing which are phenomenologically viable, $b_1$ is predominantly
the original SM bottom quark fields, $b_2$ is mostly $\omega$ and $b_3$ is mostly $\xi$.
The eigenvalues
$m_1 \equiv m_b$, $m_2$, and $m_3$ are the bottom quark mass, and two exotic
quark masses, respectively. Note that we do not necessarily order the exotic quarks $b_{2,3}$ by mass.

\subsection{Higgs Couplings}
	
The Higgs couplings are complicated by the fact that the mass matrix receives
contributions from the vector-like masses $M_2$ and $M_3$, resulting in
flavor-violating Higgs couplings between the three mass eigenstate quarks,
\bea
{\mathcal L}_{hq} = -\frac{h}{v} ~ \mathbf{\overline{d}} ~\mathbf{V_{d}} P_{R} ~\mathbf{d}
		+ h.c.
\eea
where $\mathbf{V_{d}} = \mathbf{U^{\dagger}_{d}} \mathbf{N_{d}} \mathbf{W_{d}}$.  
The off diagonal entries of $\mathbf{V_{d}}$ will lead to tree level flavor changing couplings
between the Higgs of the form $h$-$\bar{b}_1$-$b_2$, etc.  Such couplings allow for
decays of the heavy quarks into a bottom quark and a Higgs, as discussed below.

\subsection{ {\boldmath $W$} and {\boldmath $Z$} couplings}

We now examine the modifications to the $W$ and $Z$ couplings coming from the 
mixing of the bottom quark with the exotics.  In the mass basis there are $W$ couplings 
of the form
\bea
{\mathcal L}_{W} & = & 
\frac{g}{\sqrt{2}} W^-_{\mu}
\left[ \bar{\chi} \gamma^{\mu} 
\left(\mathbf{U^{2 j}_{d}} P_L + \mathbf{W^{2 j}_{d}}P_R \right) \mathbf{d^{j}} + 
\mathbf{\bar{d}^i} \gamma^{\mu} \mathbf{U^{1i*}_{d}} P_{L} t 
\right] + h.c.
\eea
where $g = e / \cos \theta_w$ as usual.

The couplings between the $Z$ and the down-type quarks may be written in matrix form,
\bea
{\mathcal L}_{Z}  & = & \frac{g}{\cos{\theta_{w}}} Z_{\mu}
\mathbf{\bar{d}} \gamma^{\mu} 
\left( \mathbf{L} P_{L} + \mathbf{R} P_{R} \right) \mathbf{d}
				+ h.c.	
\eea
where 
\bea
\mathbf{L} & = &\mathbf{U^{\dagger}_{d}} ~ \mathbf{g_{L}} ~ \mathbf{U_{d}},  \\
\mathbf{R} & = & \mathbf{W^{\dagger}_{d}} ~ \mathbf{g_{R}} ~ \mathbf{W_{d}},
\eea
and the $\mathbf{g_{L,R}}$ are diagonal matrices in the gauge basis with left and right-handed couplings of the down-type quarks to the $Z$ boson as their entries,
\bea
\mathbf{g_{L}} & = &
{\rm Diag} \left( -\frac{1}{2} + \frac{1}{3} \sin^2 \theta_w, ~~~\frac{1}{2} + \frac{1}{3} \sin^2 \theta_w,
 ~~~\frac{1}{3} \sin^2 \theta_w \right), \\
\mathbf{g_{R}} & = &
{\rm Diag} \left( \frac{1}{3} \sin^2 \theta_w, ~~~ \frac{1}{2} + \frac{1}{3} \sin^2 \theta_w,
~~~ \frac{1}{3} \sin^2 \theta_w \right).
\eea

Our primary concern is to modify the $b$-quark couplings wth the $Z$, in order to explain
the measured $A_{FB}^b$ while remaining consistent with $R_b$.  
These couplings are determined by the $11$ entries of the $\mathbf{L}$ and $\mathbf{R}$
matrices.  In terms of the parameterization, Eq.~(\ref{eq:parameterization}), these entries
are,
\bea
\mathbf{L}^{11}  &=& \mathbf{g_{L}}^{11}
\left(c^{L}_{12}c^{L}_{13} \right)^{2} 
+  \mathbf{g_{L}}^{22} \left(-s^{L}_{12}c^{L}_{23} - s^{L}_{13}s^{L}_{23}c^{L}_{12} \right)^{2} 
+ \mathbf{g_{L}}^{33} 
\left(s^{L}_{12}s^{L}_{23} - s^{L}_{13}c^{L}_{23}c^{L}_{12} \right)^{2}, \nonumber \\
\mathbf{R}^{11} &=&  
\mathbf{g_{R}}^{11} \left(c^{R}_{12}c^{R}_{13} \right)^{2} 
+ \mathbf{g_{R}}^{22} \left(-s^{R}_{12}c^{R}_{23} - s^{R}_{13}s^{R}_{23}c^{R}_{12} \right)^{2} 
+ \mathbf{g_{R}}^{33}  \left(s^{R}_{12}s^{R}_{23} - s^{R}_{13}c^{R}_{23}c^{R}_{12} \right)^{2}.
\eea
These expressions may be simplified by noting that the term proportional to the electric charge
is common to all of the diagonal entries of $\mathbf{g_{L,R}}$ and thus cancels out of shifts in
the coupling, leaving behind only the non-universal terms proportional to $T_3$.  In terms
of the mixing angles, these shifts become,
\bea
\delta g_L^b & = & \frac{g}{2 \cos \theta_w} 
\left[1 - (c^{L}_{12}c^{L}_{13})^{2} + (s^{L}_{12}c^{L}_{23} + s^{L}_{13}s^{L}_{23}c^{L}_{12})^{2}
 \right], \nonumber \\
\delta g_R^b & = & \frac{g}{2 \cos \theta_w} 
\left(s^{R}_{12}c^{R}_{23} + s^{R}_{13}s^{R}_{23}c^{R}_{12} \right)^{2}.
\label{eq:shift}
\eea

\subsection{Sample Parameters}
\label{sec:parameters}

Comparing the expressions for the shifts in the $Z$-$b$-$\bar{b}$ interactions in 
Equation~(\ref{eq:shift}) with the results in Figure~\ref{fig:deltag},
we can determine relations among the input parameters which will
improve the agreement of $A_{FB}^b$ with its measured value.

We will analyze a specific point within this parameter space and examine the collider phenomenology.  
For simplicity we assume negligible mixing between $\omega$ and $\xi$ with $s^{R}_{12} = 0.21$ and $s^{L}_{13} = 0.078$ 
and all other mixing angles set to zero. For greater clarity of notation, therefore, we denote these angles simply by $s_R$ and $s_L$ henceforth. 
The negligible mixing between $\omega$ and $\xi$ means we have chosen $Y_5 = 0$. The Yukawa couplings 
 in the bottom sector mass matrix can be related to the mixing angles in the limit of negligible $m_b$ as follows
\bea
y_L \simeq \frac{M_3 s_L}{v}
~~~~~~~~~~~~y_R \simeq \frac{M_2 s_R}{v}
\eea

The mixings which are postulated here yield couplings shifts through equation~(\ref{eq:shift}) of 

\bea
\delta g_L^b = 2.27\times10^{-3}
~~~~~~~~~~~~\delta g_R^b = 1.64\times10^{-2}
\eea

and were chosen with the intent of simplifying our discussion by removing the mixings not relevant to the measured asymmetry
while giving coupling shifts near the center of the favored region.

These assumptions leave us with only three free parameters in our system, $m_{b_2}$, $m_{b_3}$, and $m_{h}$.
The relation between $m_{b_2}$ and $m_\chi$ is

\be
m_\chi^2=m_{b_2}^2-Y_R^2=m_{b_2}c_R^2
\ee

As mentioned in \cite{Choudhury:2001hs} the fit to data is not very sensitive to $m_{b_3}$ as long as it is below a few TeV.
Looking at the  $m_h - m_{\chi}$ parameter space plot, in the study just mentioned, we choose $m_h = 120$ GeV and  
let the masses $m_{b_2}$(or equivalently $m_{\chi}$) and $m_{b_3}$ vary between 500 GeV and 2 TeV. These points lie
within 1$\sigma$ of the best fit point. We study the detailed phenomenology at this point in parameter space,
but will note where interesting deviations are possible. 


\section{Mirror Quark Phenomenology at the LHC}
\label{sec:pheno}

The key question for the LHC is whether or not the mirror quarks can be discovered, and their
$SU(2) \times U(1)$ gauge representations and mixing angles understood well enough
to experimentally connect them to the measured value of $A_{FB}^b$.  This task is complicated
by the fact that the mixing through EWSB itself obscures the original representations
of $b_2$ and $b_3$, and the $\chi$, while unusual in that it has charge $-4/3$, decays into
$W^- b$, looking much like a $t^\prime$ which produces a ``wrong sign" bottom quark in
its decay; measuring the charge of the final state $b$ quark is extremely
subtle, though perhaps not impossible \cite{Abazov:2006vd}.

\begin{figure}
\includegraphics[angle=0,scale=.75]{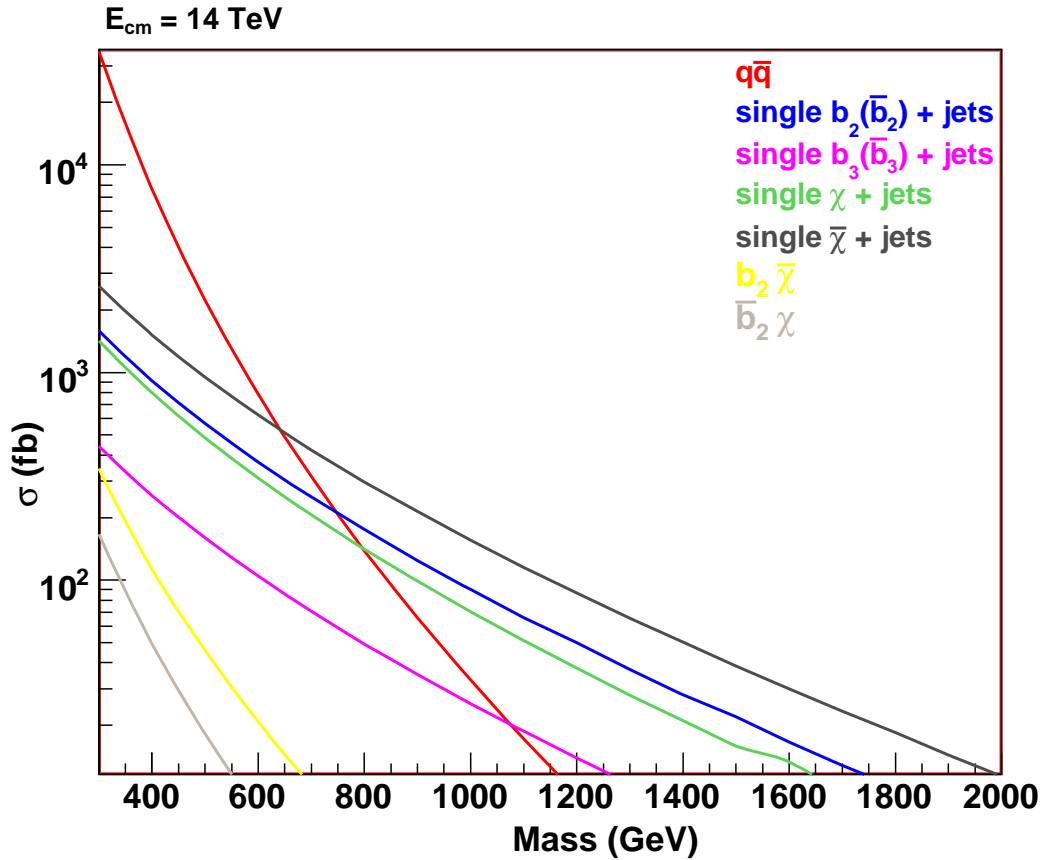} 
\caption{\label{fig:cxns}Production cross sections for both single and pair production of mirror
quarks, as a function of their masses and for the mixing angles specified in the text.  
The single $b_{2,3}$ + jets rates sum both production of $b_{2,3}$ + jets
and $\bar{b}_{2,3}$ + jets.  For this plot we have chosen $m_h = 120$ GeV}
\end{figure}

\subsection{Mirror Quark Production and Decay}

The mirror quarks $\chi$, $b_2$, and $b_3$ can be produced either in pairs through QCD, 
or singly, through the electroweak interaction.  Single $\chi$ quarks are produced through
a $b q$ initial state with a $t$-channel $W$ boson exchanged
whereas single $b_2$ and $b_3$ arise from a $b q$ initial state with a $t$-channel $Z$ boson
(or, to a much smaller degree, Higgs boson) exchanged.
The resulting cross sections as a function of 
the mass of the exotic quark in question are plotted in Fig.~\ref{fig:cxns}, where we have
used the mixing angles appropriate for the sample solution to the $A_{FB}^b$ puzzle discussed
in Section~\ref{sec:parameters} and a Higgs mass of 120 GeV (although the results are
quite robust for larger Higgs masses as well).  
The cross sections have been computed at tree level with the MadEvent code \cite{Alwall:2007st},
using the CTEQ6L parton distribution functions (PDFs) \cite{Pumplin:2002vw}.
As can be seen, for the modest mixing angles favored by $A_{FB}^b$, 
pair production is the dominant mechanism for exotic quark masses below $\sim 700$ GeV.  
The difference in rates between single $\chi$ ($Q =  -\frac{4}{3}$) and single
 $\bar{\chi}$ ($Q = +\frac{4}{3}$) production can be understood from the difference in 
 PDFs of the initial state quarks.   The $\bar{\chi}$, which is 
 primarily produced from an initial state $u$-quark, is expected to have a higher 
electroweak production rate than $\chi$ which which comes primarily from an initial $d$-quark. 
The same trend is familiar from single top production in the Standard Model.

Pair production cross sections are not affected by the choice of mixing angles at all, and thus
the prediction of that cross section is robust for any point in the parameter space of the exotic
mirrors model. Single production cross sections are proportional to the square of the relevant
mixing angle or combination of mixing angles, and thus those cross sections will be shifted by
changes in mixing angles.

The $\chi$ quark decays with $100\%$ branching ratio into $W^- b$ for our parameter point, appearing as a a 
$\bar{t}^\prime$ which produces $b$ instead of a $\bar{b}$ when it decays.  As such, it is 
sensitive to the usual fourth generation $t^\prime$ searches at the  LHC, with an
expected reach through pair production of roughly 800 GeV \cite{AguilarSaavedra:2009es}
for $100~{\rm fb}^{-1}$ of integrated luminosity at 14 TeV.
Searches for $t^\prime$ quarks at the Tevatron provide bounds of 
$m_{\chi} \geq 335$~GeV \cite{Lister:2008is}. Allowing $\omega-\xi$ mixing can open additional
channels for $\chi$ decay such as $\chi\rightarrow W^- b_3$, subject to kinematic constraints.
Depending on the size of the mixing this can become comparable in magnitude to the direct decay to
purely SM final states.

The bottom-like quarks $b_2$ and $b_3$ can decay into $Z b_i$, $h b_i$ (provided the Higgs is
light enough), $W t$, and $W \chi$.  For our example parameter point, the $\chi$ is too
heavy to be produced on-shell in decays of $b_2$, and $b_3$ does not have a charged-current
coupling to $\chi$ due to our choice of no $\omega - \xi$ mixing.
Heavy quark decays into $W t$ are rendered negligible by this assumption as well.
Thus, these quarks decay only through the FCNC modes,
\bea
b_{2,3} & \rightarrow & h b, \nonumber \\
b_{2,3} & \rightarrow & Z b. \nonumber
\eea
For Higgs mass of 120~GeV and exotic quark mass of 500~GeV, the branching ratios for both
$b_2$ and $b_3$ are  $52\%$ $b_{2,3} \rightarrow Z b$ and $48\%$ 
$b_{2,3} \rightarrow h b$. Note that these branching ratios are insensitive to changes
in the mixing angles $s_{L,R}$, and are not strongly sensitive to increases in exotic quark mass.
The $Z$ decay mode offers the possibility of lepton pairs in the
final state (with modest branching ratio)
whereas the $h$ decay mode leads to a $b_{2,3} \rightarrow b \bar{b} b$
final state a large fraction of the time.

Branching ratios of $b_i$ decays are independent of shifts in $s_L$ and $s_R$ in absence of $\omega-\xi$
mixing. This is because in that limit the flavor changing couplings of the higgs and the $Z$ both
have identical dependence on the mixing angles. If mixing between the new vector-like quarks is
allowed it can lead to shifts in the relative branching fractions to $Z$ and $h$ final states, and also
opens additional decay channels, such as $b_3\rightarrow W+ \chi$ or $b_2\rightarrow Z b_3$,
dependent on kinematic constraints. These will in general lead to more spectacular cascade decays,
as the decay product vector quark then further decays to SM fields.

While the LHC cannot hope to exclude the entire range of mirror quark masses favored by the
electroweak fit, it is sensitive to much of the parameter space.  For masses $\lesssim 1$~TeV,
we can expect based on earlier studies \cite{AguilarSaavedra:2009es}
that by the end-running of the LHC (which we take to be at center-of-mass
running of 14~TeV and data sets on the order of $300~{\rm fb}^{-1}$), the LHC will
have observed the mirror quarks through pair production in the decay modes
$\chi \rightarrow W^- b$ and $b_{2,3} \rightarrow Z b \rightarrow \ell^+ \ell^- b$.  We explore 
several subdominant production processes which can help differentiate the beautiful mirrors
model from other models with additional vector-like quarks, and establish the mixing
parameters as consistent with a solution to $A_{FB}^b$.  We choose as a reference
value for our studies mirror quark masses $M_2 ( = m_\chi) = 500$~GeV.  Such masses are
consistent with Tevatron bounds and within the $1\sigma$ fit to the precision data
\cite{Choudhury:2001hs}, and represent a cautiously optimistic 
 region of parameter space.  We assume $M_3$, which is not well constrained
by the fit, is $\geq 1 $~TeV, and thus do not assume $b_3$ will be observable.

\subsection{ {\boldmath $\bar{\chi}$} Production}

The process $u \bar{b} \rightarrow d \bar{\chi}$ is the largest of the single production
modes in the model, and under our assumption that $M_3 \gg M_2$, its rate
is proportional to $s^2_R$, thus providing a measure of the key mixing
which is responsible for $\delta g_R$.  There is also a contribution from the left-handed
mixing, but this is constrained to be small by precision data.  We attempt to extract the
signal (and thus measure $s_R$) by looking at the semi-leptonic $\chi$
decay:  $pp \rightarrow j \bar{\chi} \rightarrow j b \ell \nu$ where $j$ is a light-quark
initiated jet and $\ell = e$ or $\mu$. To improve background rejection, we do not attept
to reconstruct single $\chi$ production here, focusing on the dominant single $\bar{\chi}$ signal.
For our sample parameter point, the signal inclusive cross section, including
branching ratios, is $949$~fb.  The SM background (with very mild acceptance
cuts on the jets) is $12.7$~nb, dominantly 
$Wjj$ production, with smaller contributions from $t \bar{t}$ and single top production.
Events are showered and hadronized with PYTHIA \cite{Sjostrand:2006za}, 
and we estimate detector 
effects with PGS \cite{pgs} using the default LHC detector model of MadEvent.

To separate the signal from the background efficiently, we require that the event contain 
exactly one $b$-tagged jet with transverse momentum $P_T\geq100$~GeV, one positively-charged lepton with 
$P_T\geq50$~GeV (which is sufficient to trigger on the events even in high luminosity running), 
no more than two jets with $P_T\geq30$~GeV, and missing momentum 
$\not \hspace*{-.1cm} E_T\geq50$~GeV. 
We further require that the invariant mass of the two highest 
$P_T$ jets $M_{jj}$ be $\geq100$~GeV.  We assume the $\not \hspace*{-.15cm} E_T$ arises
from a neutrino present in a on-shell
$W$ decay, and use the $W$ mass to reconstruct the longitudinal
neutrino momentum.  Armed with that information, we can reconstruct the four-momentum 
for the $W$ boson, which we combine with the $b$-tagged jet to form the invariant mass which
in a signal event would reconstruct the $\chi$ mass of 500~GeV.  We apply a wide cut to
this quantity, requiring it to be in the range $400-600$~GeV.

We found a signal acceptance of about $1\%$ and
a background suppression factor of $2.6\times10^{-5}$ using these cuts.
While the signal-to-background 
ratio remains small, sufficient statistics can be generated for a significant observation
of the process and a measurement of the signal cross section. 
With $100\ {\rm fb}^{-1}$ of data the total number of expected events exceeds the SM prediction by 
950, equivalent to $5.2\sigma$, constituting a discovery of the single production process for 
$\bar{\chi}$, and with $300\ {\rm fb}^{-1}$ the cross section can be measured to be 
$949\pm147\ {\rm fb}$, where systematic uncertainties in determining the acceptances are 
assumed to be small in comparison to the large statistical uncertainties. This measurement
corresponds to 2,800 expected signal events over 99,000 expected background events.
Extracting $s_R$ from the cross section is straightforward, as all other quantities
entering the conversion are known with effectively zero error compared to the measured cross
section.  We end up with a measurement of $s^2_R=0.044\pm0.007$.

The discovery potential in this channel extends well beyond the mass studied here. Azuelos {\it et al}
\cite{Azuelos} found that a vector like $t^\prime$ was discoverable through
the $t^\prime\rightarrow W b$ channel up to $m_{t^\prime}\simeq 2.5$~TeV. Our model predicts
an identical signal, with comparable production cross section and more favorable branching
ratio for this measurement.

The fact that $\chi$ has charge $-4/3$ is a very distinctive feature compared to other
models of vector-like quarks, but difficult to establish experimentally.  One could attempt
to measure the charge of the $b$ quark produced in a $\chi$ decay; this has been
successfully employed by Tevatron experiments to establish the top quark
charge \cite{Abazov:2006vd}, but depends sensitively on modeling the detector response
correctly, and thus is beyond the scope of this work.  Additional strategies could
be to examine processes such as $\chi \bar{\chi} \gamma$, which is expected to lead
to a successful LHC measurement of the top quark charge \cite{Baur:2001si}.  We have
performed simulations of $\chi \bar{\chi} \gamma$ production, 
but find that the contribution induced by photon radiation from
the parent quark becomes lost in radiation from the $W$ or lepton in its decay.  The
large $\chi$ mass has the unfortunate effect of both reducing the over-all rate substantially
compared to the $t \bar{t} \gamma$, and also collimates the $\chi$ decay products, making
it more difficult to extract the cases where the photon is radiated by the final state lepton from
that where it is radiated from the quark itself than was true for the well-spread out top
quark decay products.

Ultimately, the most promising argument for the charge of the $\chi$ may be indirect by
the failure to observe the decay mode $\chi \rightarrow Z t$, which would generically
be present for a charge $2/3$ vector quark, which would be allowed to mix with the top.
This argument rests on the assumption that 
one has observed $b_2 \rightarrow Z b$, and thus knows that the newly discovered objects
are in fact vector-like as opposed to chiral quarks.  However, it is worth bearing in mind
that even for a vector-like $t^\prime$, the $Z$-$t^\prime$-$t$ interaction is controlled by
separate mixing angles from those in the $b$ sector, and thus
may turn out to be very small\footnote{In fact, the reasonable agreement between the
experimental measurements of $b \rightarrow s \gamma$ and SM predictions
{\em requires} that the product of the $t$-$t^\prime$ and $b$-$b^\prime$ mixings 
be $\lesssim 10^{-2}$ \cite{Larios:1999au}.}.

\subsection{Single {\boldmath $b_{2}/\bar{b_{2}}$} production}

After single $\bar{\chi}$ production, the next largest single mirror quark production mode
is single $b_2$ production (including single $\bar{b}_2$ production), which
proceeds through an FCNC $Z$ or $h$ exchange in the $t$-channel.  The rate for this process
is proportional to $s^2_R c^2_R$, and thus
provides another measurement of the mixing angle $s^2_R$.
We examine the feasibility of observing the process 
$pp \rightarrow j b_2 \rightarrow j b \ell^+ \ell^-$  through an intermediate $Z$ boson
from the $b_2$ decay
(and also the conjugate process for $\bar{b}_2$).
The signal cross section (including branching ratios) for
$m_{b_2} = 500$~GeV is 16.6 fb.  The background is dominantly $Z b \bar{b}$ and
$t \bar{t}$ and is 125 pb after acceptance cuts.

We require at least one of the leptons in the event  to have $P_{T} \geq$ 20 GeV, sufficient for
triggering.  We select events with at least one $b$-tagged jet with $P_{T}\geq$ 70 GeV.
The main criteria to distinguish between the signal and background are the reconstructed mass 
of two leptons, which should be close to the $Z$ mass and its combination with the 
$b$-jet to form an invariant mass close to $m_{2}$.  In events with more than one $b$-tagged 
jet (as is often the case for the background processes), we combine the $b$-jet that has the 
largest $P_T$ with the lepton pair to form the reconstructed $b_2$ mass.  We require the 
reconstructed $b_2$ mass to be within a 25 GeV window of the 
reference value of $m_2 = 500$~GeV.  This window contains $42.3\%$ of the signal rate
after jet smearing.
In addition, we place a restriction on the angular separation of $\Delta R \leq 1$
between the pair of leptons, since the signal produces highly boosted $Z$ bosons from the
$b_2$ decay whose decay products are collimated. 

At an integrated luminosity of $300~{\rm fb}^{-1}$ the number of predicted signal events
for our parameter point is 280, after applying the above mentioned cuts. The analysis
efficiency is $5.6\%$. The number of expected background events is 1260 after background
suppression by a factor of $3.4\times10^{-5}$. The
resulting significance is 7.9$\sigma$, and the measurement of the mixing
angle is $s^2_R=0.044\pm0.008$, which is comparable to the precision offered
by single $\bar{\chi}$ production.  While not a direct reconstruction of the vector-like
quarks, these two measurements together provide evidence that the primary mixing
with the third generation SM quarks is through the bottom sector, with no apparent
mixing involving the top (which is forbidden in our construction by $U(1)_{EM}$ but
could be allowed in generic models of mirror quarks containing top-like objects).

\subsection{Electroweak {\boldmath \bf{$b_2-\chi$}} production}

Another process which allows us to extract information about the mixing angle is electroweak
production of a pair of mirror quarks, $\chi b_2$ through an $s$-channel $W$ boson.  
The cross section, which is proportional to $c_R^2$, turns out to be quite small due to the fact 
that on top of being governed by weak couplings, two heavy quarks are being produced.  
To analyze this signal we look at the process 
$pp \rightarrow \chi b_2 \rightarrow \ell^{\pm} \ell^{\mp} b\bar{b}  \ell^\prime \nu$ 
(and its charge-conjugate version).  This particular signature has the $b_2$ decaying 
through a $Z$ into $\ell^+ \ell^-$ whereas $\chi$ decays as usual into $W b$.
After acceptance cuts, the cross section for this signal is $0.359$ fb.  The background
was generated using MadEvent and the relevant decays were obtained with BRIDGE \cite{bridge}
before showering and hadronizing with PYTHIA.  Again detector effects were estimated with PGS
using the default LHC detector model of MadEvent.  This resulted in a background cross section of $3.8$ fb.

Due to the small number of events, distinguishing signal from background is difficult.  In order to 
retain enough events to obtain sufficient statistics one must be conservative in applying cuts.  
We first require that the event contain two $b$-tagged jets with  $P_{T}\geq$ 60 GeV.  Since we
expect the charged lepton pair decaying from the highly boosted Z to be collimated we first find
the  $\ell^{\pm} \ell^{\mp}$ pair with the smallest  $\Delta R$ and combine it with one of the
$b$ jets to form an invariant mass (in a signal event this would reconstruct the $b_2$ mass)
which we require to be greater than 50 GeV.  

To determine which lepton is associated with the neutrino we again assume the $\not \hspace*{-.15cm} E_T$
arises from a neutrino present in an on-shell $W$ decay.  We then find any charged leptons not belonging
to the pair which decay from the Z and out of those find the one with the smallest $\Delta \phi$ relative
to the missing $E_T$.   We use the $W$ mass to reconstruct the longitudinal neutrino momentum, and
from that we reconstruct the four-momentum for the $W$ boson and combine it with the other $b$-tagged
jet with  $P_{T}\geq$ 60 GeV to obtain the invariant mass  (in a signal event this would reconstruct the
$\chi$ mass) which we require to be greater than 200 GeV.

With a signal acceptance of $6.25\%$ these selection criteria lead to an expected 10 signal events at
$300~{\rm fb}^{-1}$.  We expect 3 background events after a suppression factor of $2.4\times 10^{-3}$
leading to a significance of $4\sigma$ for the signal over background.  While this does not constitute
a discovery, when combined with the information obtained from other signals, it provides evidence that
the $\chi$ and $b_2$ quarks form an $SU(2)$ doublet, and thus does help to verify the structure of the
Beautiful mirrors model.

\section{Conclusions}
\label{sec:conclusion}

While the possibilities for discovery at the LHC are vast, it may be that there are clues as to what
could be discovered in the form of modest deviations already present in lower energy data.
In this article, we have discussed one such deviation, the forward-backward asymmetry of the
bottom quark, which has persisted for more than a decade and appears to play a key role in the
SM fit to the Higgs mass.  We have explored one particular model which addresses the 
discrepancy by adding additional vector-like quarks which mix with the $b$, subtly affecting
its coupling to the $Z$ boson.

These quarks are perfect targets for discovery at the LHC, which is likely to initially observe
them through pair production.  We have examined the prospects for observing single production
as well.  While single production has smaller rates, being suppressed by electroweak strength
couplings and mixing angles, it probes the basic phenomena responsible for the solution to
the mystery of $A_{FB}^b$.  In particular, we have studied single $\bar{\chi}$ production followed
by the dominant decay $\bar{\chi} \rightarrow W^+ \bar{b}$, single $b_2$ production followed by the
decay $b_2 \rightarrow Z b \rightarrow \ell^+ \ell^- b$, and pair production of $\chi \bar{b}_2$
(with the same decay chains as above).  These processes are expected to be visible for
quark masses up to about 500 GeV
at a 14 TeV LHC with hundreds of ${\rm fb}^{-1}$, and provide evidence that the
$\chi$ and $b_2$ quarks form a vector-like electroweak doublet which mixes primarily
with the bottom quark.  The primary mixing parameter, $s_R^2$, responsible for explaining the value
of $A_{FB}^b$, can be measured at the $20\%$ level.

\acknowledgements

The authors are grateful for conversations with Carlos E.M. Wagner.  
T Tait appreciates the SLAC
theory group for their extraordinary generosity during his many visits.  
KK and R V-M would like to thank Gabe Shaughnessy, Jamie Gainer, and Patrick Fox for helpful discussions.
Research at Argonne National Laboratory is 
supported in part by the Department of Energy under contract
DE-AC02-06CH11357.


\end{document}